\documentclass{article}

\usepackage{arxiv}

\usepackage[utf8]{inputenc} 
\usepackage[T1]{fontenc}    
\usepackage{hyperref}       
\usepackage{url}            
\usepackage{booktabs}       
\usepackage{amsfonts}       
\usepackage{nicefrac}       
\usepackage{microtype}      
\usepackage{lipsum}
\usepackage{amsmath}
\usepackage{amsthm}
\usepackage{amssymb}
\usepackage{booktabs}
\usepackage{multirow}
\usepackage{multicol}
\usepackage{graphicx}
\graphicspath{ {./images/} }

\title{Predicting  Distance Matrix with Large Language Models}

\author{
 Jiaxing Yang \\
 Shanghai Artificial Intelligence Laboratory\\
Shanghai, Shanghai 201100, China\\
\texttt{jxyang900\@gmail.com} \\
}

\begin{document}
\maketitle
\begin{abstract}
Structural prediction has long been considered the crown jewel of RNA research, especially following the success of AlphaFold2 in protein studies, which has drawn significant attention to the field. While recent advances in machine learning and data accumulation have effectively addressed many biological tasks—particularly in protein-related research—RNA structure prediction remains a significant challenge due to data limitations. Obtaining RNA structural data is difficult because traditional methods such as nuclear magnetic resonance spectroscopy, X-ray crystallography, and electron microscopy are expensive and time-consuming. Although several RNA 3D structure prediction methods have been proposed, their accuracy is still limited. Predicting RNA structural information at another level, such as distance maps, remains highly valuable. Distance maps provide a simplified representation of spatial constraints between nucleotides, capturing essential relationships without requiring a full 3D model. This intermediate level of structural information can guide more accurate 3D modeling and is computationally less intensive, making it a useful tool for improving structural predictions. Therefore, to provide more accurate predictions of RNA 3D structures from a machine learning perspective, we are the first to explore the task of defining the distances between arbitrary base pairs in the RNA primary sequence. This regression task offers more informative data for subsequent 3D folding methods but is more complex than the well-known RNA secondary structure prediction. In this work, we demonstrate that using only primary sequence information, we can accurately infer the distances between RNA bases by utilizing a large pre-trained RNA language model coupled with a well-trained downstream transformer. Furthermore, by converting our distance prediction outputs into contact predictions, we achieved performance comparable to other methods at the contact prediction level.
\end{abstract}


\section{Introduction}

RNA structures are foundational for understanding their diverse cellular functions, such as cell signaling~\cite{geisler2013rna}, gene expression regulation~\cite{statello2021gene}, and post-transcriptional regulation~\cite{alipanahi2015predicting}. Additionally, determining RNA structure is essential for RNA-based therapeutics, including mRNA vaccines~\cite{pardi2018mrna}, RNA interference~\cite{bora2012rna}, and CRISPR-based therapies~\cite{li2020strategies}. Traditionally, RNA three-dimensional (3D) structures are assessed using experimental approaches such as nuclear magnetic resonance (NMR) spectroscopy, X-ray crystallography, and cryogenic electron microscopy. However, these methods are expensive and time-consuming. Compared to proteins, the number of experimentally validated RNA structures is low, accounting for only 3\% of RNA families in Rfam due to their biochemical instability~\cite{kalvari2018rfam}. Consequently, there is a pressing need to interpret RNA sequences into structures using \emph{in silico} methods.

To bridge the gap between RNA sequences and their 3D structures, many secondary structure prediction approaches have been developed, such as E2Efold~\cite{chen2020rna}, CONTRAfold~\cite{do2006contrafold}, and RNAfold~\cite{gruber2008vienna}. While these methods provide valuable insights into base pairing interactions, they do not capture the full 3D conformation of RNA molecules. To address this limitation, contact maps have been introduced to represent base interactions in RNA structures, as seen in methods like SPOT-RNA~\cite{singh2019rna}, RNAmap2D~\cite{pietal2012rnamap2d}, and DIRECT~\cite{jian2019direct}. Contact maps offer a simplified representation of structural information by indicating whether pairs of nucleotides are in contact, typically within a certain distance threshold. 

Despite these advances, predicting the precise spatial relationships between nucleotides remains a significant challenge. Although several RNA 3D structure prediction methods have been proposed, their accuracy is still limited~\cite{miao2017rna,shen2022e2efold,li2023integrating}. Predicting RNA structural information at another level, such as distance maps, remains highly valuable. Distance maps provide detailed spatial constraints between nucleotides, capturing essential relationships without requiring a full 3D model. This intermediate level of structural information can guide more accurate 3D modeling and is computationally less intensive, making it a useful tool for improving structural predictions.

Here, we propose a novel approach to tackle the RNA distance prediction problem by examining the Euclidean distances between arbitrary bases in the RNA primary sequence to assess RNA 3D structure. Intuitively, a distance matrix takes a step closer than a contact map to represent the 3D structure, providing detailed spatial constraints that are essential for accurate 3D modeling. However, RNA distance matrices have been seldom considered due to the limited availability of comprehensive RNA structural data. In protein research, distance matrices are regarded as simplified representations of protein structures and have been widely used in \emph{ab initio} protein structure prediction~\cite{chen2019improve}. This work proposes a new method that predicts the RNA distance matrix directly from its sequence by leveraging pre-trained RNA language models. Unlike traditional convolutional-based algorithms commonly used in protein research, we argue that the attention mechanism inherent in transformers naturally aligns with the task of predicting nucleotide pair distances. The attention mechanism can capture long-range dependencies and complex interactions between nucleotides, which are crucial for accurate distance predictions. Therefore, we adopt the vanilla transformer architecture and name our framework the Distance Transformer (DiT). Our proposed method represents another view forward in RNA structure prediction. By predicting the RNA distance matrix from sequence data, we can enhance the understanding of RNA structures and their functions, facilitating advancements in both basic research and therapeutic applications.

\section{Methods}
Distance in RNA research refers to the  euclidean distance between atoms C$\alpha$-C$\alpha$  among every pair of nucleotide base. A length $n$  RNA nucleotide sequence, therefore 
contains $n\times n$ base pairs and could be presented as a two-dimensional distance matrix. 

\begin{figure*}[h]
    \begin{center}
    \includegraphics[width=0.8\textwidth]{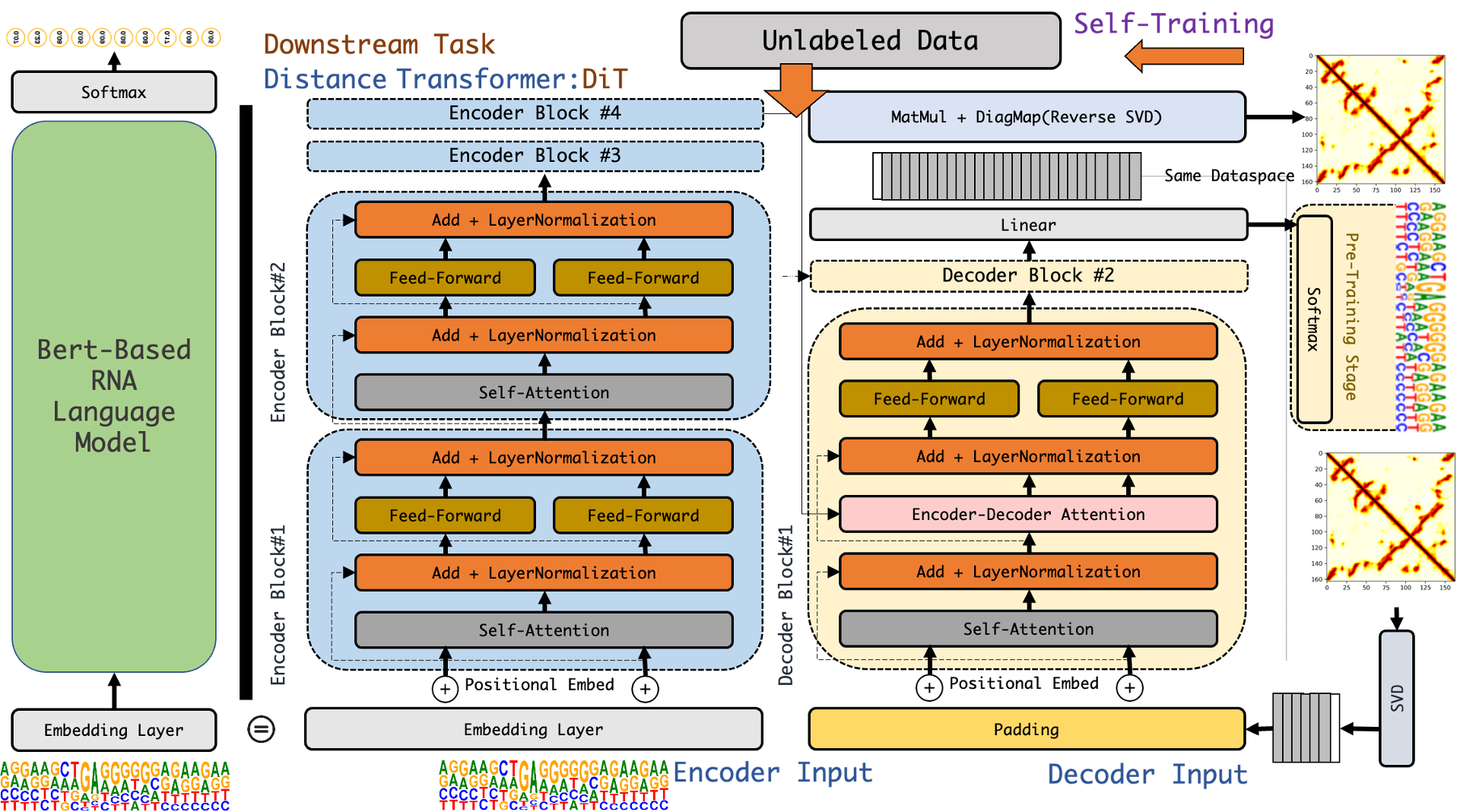}
    \caption{Overview of our whole framework. The large scale RNA Language Model phase 
    served as a ‘green box’ providing us with trained embedding layer. Right half presents 
    the DiT architecture pre/self-training stage.}\label{Trainig Stage}
    \label{fig:Fig1}
    \end{center}
    
\end{figure*}

Our task thus could be formulated as predicting a two dimensional $X$ which is non-negative, symmetric and zeroed on diagonal,
\begin{eqnarray}
  X\in S^{n} \:,\:  x_{ii} = 0, 1\leq i \leq n \nonumber  
  \end{eqnarray}
 In this work, we will be predicting the distance matrix from scratch by only using the sequential input, no other auxiliary information is required (such as Secondary-Structure, Multiple Sequence Alignment).
 ‘The less, the better’, this critical advantage of our framework enables our framework to scale nicely in various scenarios. 
The whole framework involves two phases where we obtained sequential embedding from a 
 RNA bidirectional language model and deployed a distance transformer (DiT) on our downstream task.
As shown in Fig.\ref{fig:Fig1}, we will treat the upstream module as a standalone ‘black box’ which
provides us with the embedding layer and directly look into our downstream design. 
 
 Here, DiT training can also be divided into multiple stages: model pre-training, distance matrix tuning and self-training. Unlike our first phase, DiT is of relatively small capacity and  our pre/self-training 
 works on a small dataset consisting of around $2k$ sequential data $1k$ of groundtruth distance matrix obtained from Protein Data Bank. 

\subsection{RNA Bidirectional Language Model}

The intention of our first phase aims to provide rich RNA sequence 
representations for further downstream tasks as prescribed. We used RNA-FM where a Bert-based 
language model with 12 transformer encoder blocks \cite{devlin2018bert,chen2022interpretable} was 
trained on around 26 million non-coding RNA sequences in an unsupervised manner. After the training stage, 
a learned embedding layer will matrix an RNA sequence of length $L$  to a $L\times 640$ tensor.

Noticed that this trained RNA Bert model could be applied 
directly for fine-tuning in other tasks concatenated with output heads. However, the difficulty 
lies in the gap between enormous pre-trained capacity 
and relatively small downstream datasets in different feature spaces. 
Thus, a distance transformer (DiT) is implemented for tackling the downstream distance prediction problem.

\subsection{Distance Transformer}
Referring to such a task, we claim that it naturally agrees to the computation of self-attention mechanism.
In fact, considering the distance between all nucleotide pairs actually conveyed to calculate the relative ‘attention’ among these pairs. Thus, different from the former convolutional design in ‘Protein’, 
an intuitive structure could be applying stacked 
attention modules on top of input nucleotide sequences.

Since transformer based models have achieved great success in both CV \cite{dosovitskiy2020image} and
NLP \cite{vaswani2017attention} tracks, here, we applied vanilla transformer architecture 
for our DiT model with fewest possible modifications to testify our above statement. Our DiT follows the 
standard transformer and consists
of an Encoder-Decoder form, each Encoder-Decoder block within is just stacks of multi-head-attention and layer normalization.

As the size of our distance training dataset stays beyond $1k$, it requires more carefully designed
 training strategies to tackle the problem. Suggested by \cite{zoph2020rethinking}, pre-training and self-training on a comparable size 
with small labeled datasets almost always improve model accuracy. Thus, our training strategy involves pre-training DiT on $2k$ sequences, tuning on a $1k$ distance/sequence dataset, and self-training 
combined with remaining $1k$ sequences, which distance data isn't available.


\subsubsection{DiT Pre-Training}

Unsupervised pre-training has been widely applied in the field \cite{vaswani2017attention}
\cite{DBLP:journals/corr/abs-2111-06377}. Similarly, we directly applied the standard language model \cite{vaswani2017attention}
to avoid changes in the model.
As shown in the right part of Fig.\ref{fig:Fig1}, 
  The input of our encoder admits a batch of nucleotide sequences, 
  input of decoder takes the same batch of nucleotide sequences with a right shift 
  of ‘Empty Token’. Outputs are produced via the ‘grey’ linear layer
  and do not flow through ‘MatMul’. Finally, softmax above the linear layer estimates probabilities
  of the next nucleotide character.

  Here, we regularized each nucleotide sequence to be of length $512$, those 
  under $512$ would be padded, and oversized would be cut off to this length. 
  Each nucleotide's feature dimension is $648$ containing $640$ of embedding dimension 
  and $8$ of one-hot encoding. A dropout rate of $0.2$ and label smoothing of $0.1$ is employed.

  \subsubsection{Distance matrix Tuning}

  After pre-training on $2k$ sequences data, we adapt DiT on $1k$ distance matrix data for tuning. 
  Here we will need to modify our input/output layer for tuning purposes. 
  Since distance matrix is symmetric and zeroed on its diagonal, we can re-design our final output layer by taking advantage of such prior knowledge.
  'MatMul + DiagMap' layer does the following calculation:

  \begin{eqnarray}
      &f(A) =  A \cdot  \Sigma(x) \cdot  A^{\intercal} \nonumber      \\ 
      &g(a_{ii})=0 \:\:\:\:\:\: g(a_{ij})  = a_{ij}, i\neq j \nonumber      \\
      &Output =  g \circ f(A)  
    \end{eqnarray}

Where $A$ stands for input from the linear layer of size $L \times 648$(
    where $L$ here denotes $512$
), $\Sigma(x)$ refers to a learnable diagonal matrix, $T$ for the transpose,
 and $g$ simply as a mask function. Noticed that $f$ here follows a 
 SVD process which generates a square symmetric matrix. Thus, eventually, 
 $g \circ f(A)$ takes the form of a distance matrix. Now, consider the reverse, 
 will any distance matrix $512\times512$ follow a decomposition of $A \cdot  \Sigma \cdot  A^\intercal$, 
 where $A$ of size $512 \times 648$?

 This is true since $\forall X \in \mathcal{S}^{n}$, 
 , there admits a decomposition $X = B \cdot  \Sigma' \cdot  B^\intercal$ \cite{bhatia2013matrix}, where 
  $B$ of size $n \times n$. Then, by applying a block transform,

  \begin{eqnarray}
        A =  \begin{pmatrix}
            B  \\ 
            0 
          \end{pmatrix}, \Sigma = \begin{pmatrix}
            \Sigma' & 0\\ 
            0 & 0
          \end{pmatrix}   
    \end{eqnarray}
    \begin{eqnarray}
            A\Sigma A^{\intercal} &=&  \begin{pmatrix}
                B  \\ 
                0 
              \end{pmatrix} \cdot \begin{pmatrix}
                \Sigma' & 0\\ 
                0 & 0
              \end{pmatrix} \cdot (B^{\intercal} \: \: 0) \nonumber  \\
                &=& \begin{pmatrix}
                    B\Sigma' & 0\\ 
                    0 & 0
                  \end{pmatrix} \cdot (B^{\intercal} \: \: 0)\nonumber \\
                  & = &  B\Sigma'B^{\intercal} \nonumber \\
                  & = &  X
        \end{eqnarray}
        \begin{figure}[]
            \begin{center}
            \includegraphics[width=0.32\textwidth]{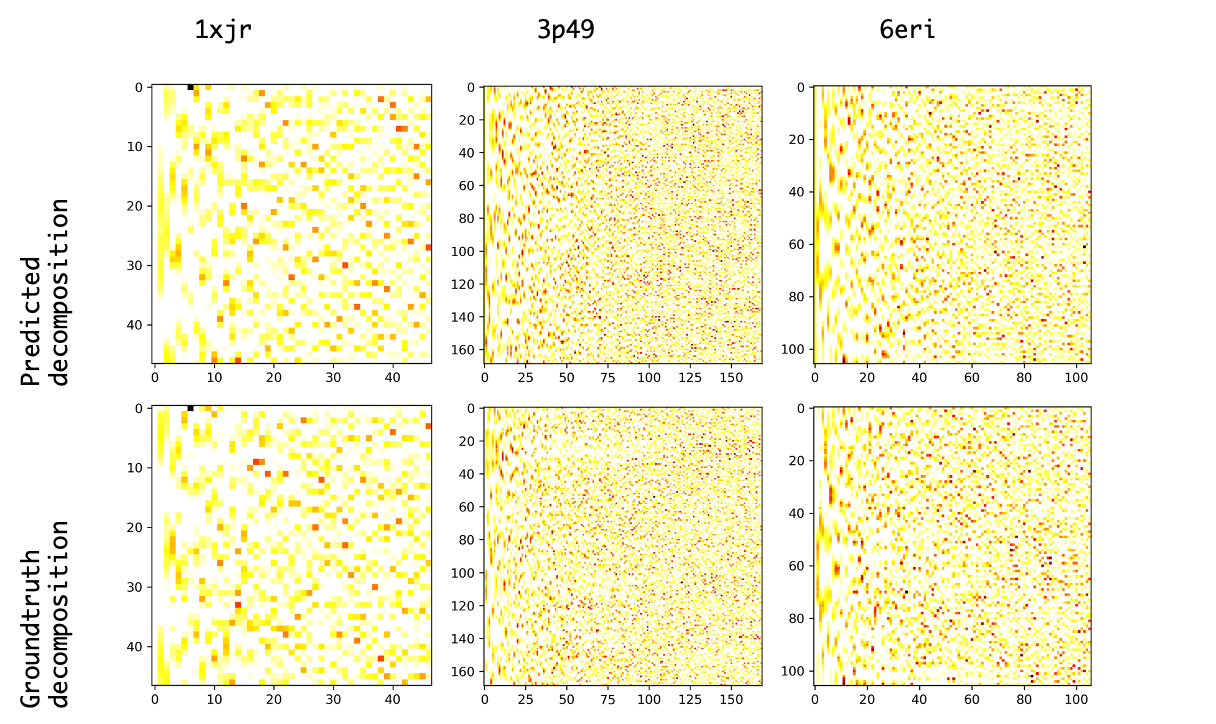}
            \caption{Illustration of what DiT is learning from distance data. The learned feature matrix is a decomposition $B$ of a distance matrix $X$. }\label{Trainig Stage}
            \end{center}
        \end{figure}
Thus, from Eq.2, for any distance matrix $X$, the input of decoder follows a
right ‘Empty Token’ shift of $A = \big(\begin{smallmatrix}
    B\\
    0 
  \end{smallmatrix}\big)$, and the output is expected to be the same tensor $A$ with a left ‘Empty Token’
  shift. As shown in the right part of Fig.\ref{fig:Fig1}, 'SVD' and ‘MatMul + DiagMap’ layer computes 
  in the above reverse scheme. Fig.2 gives some ideas about how $B$ looks
   like and what does our model learns.

   \subsubsection{DiT Self-Training}

   After tuning done on our $1k$ distance data, we reach the final stage of 
   self-training on remaining unlabeled $1k$ data to boost our performance further. 
   Training details are shown in Fig.3; We generate pseudo distance matrix of the
    unlabeled data and select a portion (starting with shorter sequences) to combine with 
    groundtruth labels. 
   
    \begin{figure}[bh]
        \begin{center}
        \includegraphics[width=0.32\textwidth]{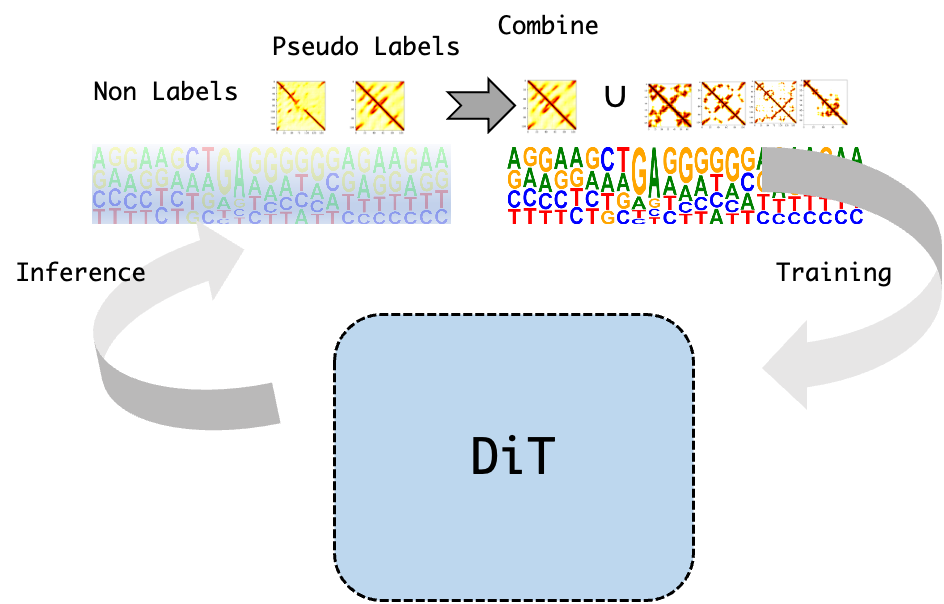}
        \caption{Self-training stage where we iteratively generates pseudo labels of unlabeled data and re-trains the model from distance tuning combining those pseudo data.}\label{dd}
        \end{center}
    \end{figure}

 Next, a new iteration of training involving ‘distance matrix tuning’ and ‘self-training’
 is initiated. Self-training is done when the process reaches max iteration or obtains no performance gain.

\section{Results}
In this section, we evaluate our model in three aspects. Firstly, we look into 
our performance when considering distance prediction as a vanilla regression problem, and since 
our work is the first to tackle the RNA distance problem, we compare DiT with traditional convolution-based
 network, for instance, Unet++ \cite{zhou2018unet++} which ranks top1 in several medical segmentation tasks.

 Secondly, since protein or RNA distance matrix serves as an essential role in defining their 3D-Structure, we study 
 from the side of 3D-Structure predictions' accuracy.

 Finally, although there's no other work on our task, we can evaluate our downgraded version of predicting RNA contacts with other methods.
 Impacts of model capacity are also done in this section.

\subsubsection*{Data Pre-Processing} The dataset of our work consists of the non-redundant RNA 3D structures \cite{leontis2012nonredundant}(Release 3.202) which is updated weekly. This dataset organizes all RNA-containing 3D structures from PDB into sequence/structure equivalence classes 
and selects a high-quality representative structure from each category. After disposing of sequences that are shorter than $30$, we obtained around $2k$ sequences. Only about $1k$ distance matrix could be generated using pydca tool \cite{zerihun2020pydca}.
Generated distance are clipped to $[0,\:20]$ and divided by $20$ ranging to $[0,\:1]$.
\begin{table}
    \centering
    
    \label{model-performance}
    \begin{tabular}{cccccccc}\toprule
     \multirow{2}{*}{Methods} & $\multirow{2}{*}{MSE}$ & $\multirow{2}{*}{MAE}$ 
    & $\multirow{2}{*}{$R^{2}$}$ & $\multirow{2}{*}{PA\%}$ &  $\multirow{2}{*}{PMCC}$  \\ \\\toprule
     \textbf{DiT-PS} & \textbf{0.021} & \textbf{0.124} & \textbf{0.821} & \textbf{92.28}  & \textbf{0.93}  \\
       DiT & 0.032 & 0.171 & 0.792 & 90.23& 0.90  \\
 Unet++ (Seq+SS) & 0.035 & 0.199 & 0.751 & 83.23 & 0.78 \\
 Unet++ (Seq) & 0.039 & 0.212 & 0.598 & 61.23 & 0.66 \\\bottomrule
\end{tabular}\caption{Distance performance Evaluation of on the independent test Dataset TE100 where we treat the task as a vanilla regression problem.}
\end{table}

\begin{figure*}[]
    \begin{center}
    \includegraphics[width=0.8\textwidth]{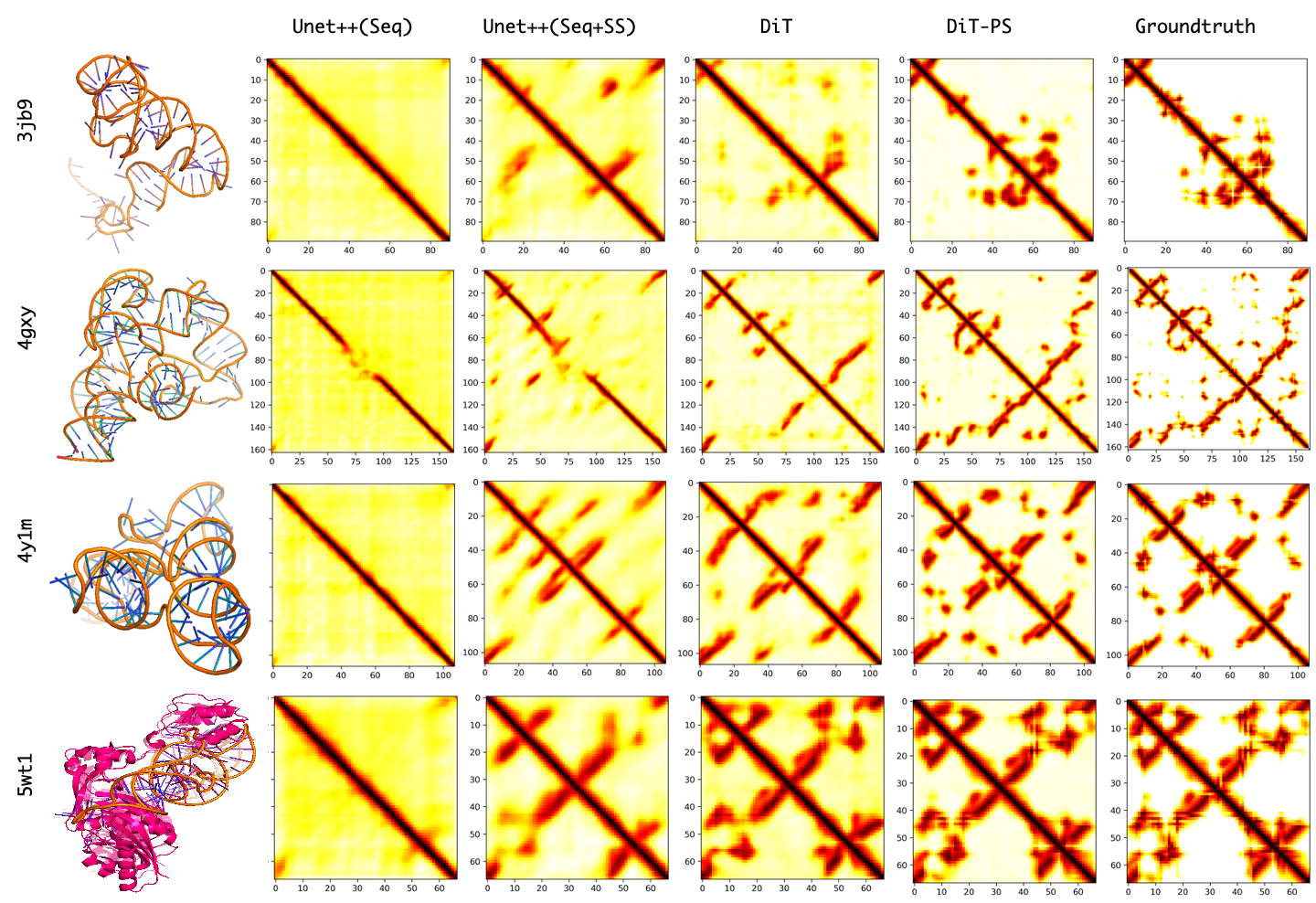}
    \caption{Visualization of predicting results on four RNA structures 5wt1,4y1m,4gxy,3jb9. DiT-PS can generate almost equivalent patterns on complicated structures.}\label{12341234t}
    \end{center}
\end{figure*}

\begin{figure*}[]
    \begin{center}
    \includegraphics[width=0.88\textwidth]{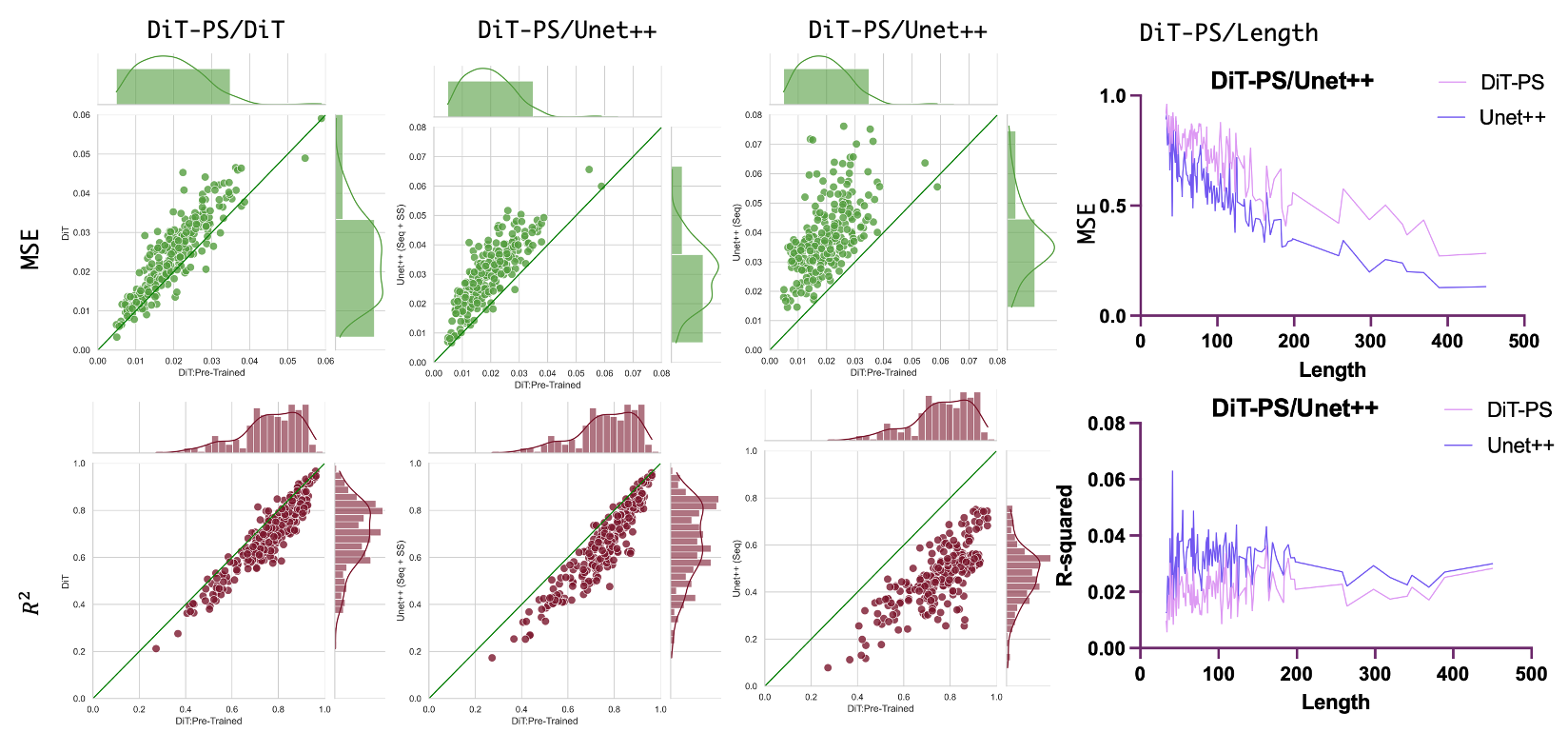}
    \caption{Left part presents the scatterplot comparison of DiT-PS to other alternatives, almost 
    all data points stayed from the diagonal reflecting we outperforms other alternatives with a preferable gap.
    Right part suggests that DiT-PS performance drops when structure complexity grows, but we're still better that convolutional-based Unet++.}\label{tasd}
    \end{center}
\end{figure*}

\begin{figure*}[h]
    \begin{center}
    \includegraphics[width=1.0\textwidth]{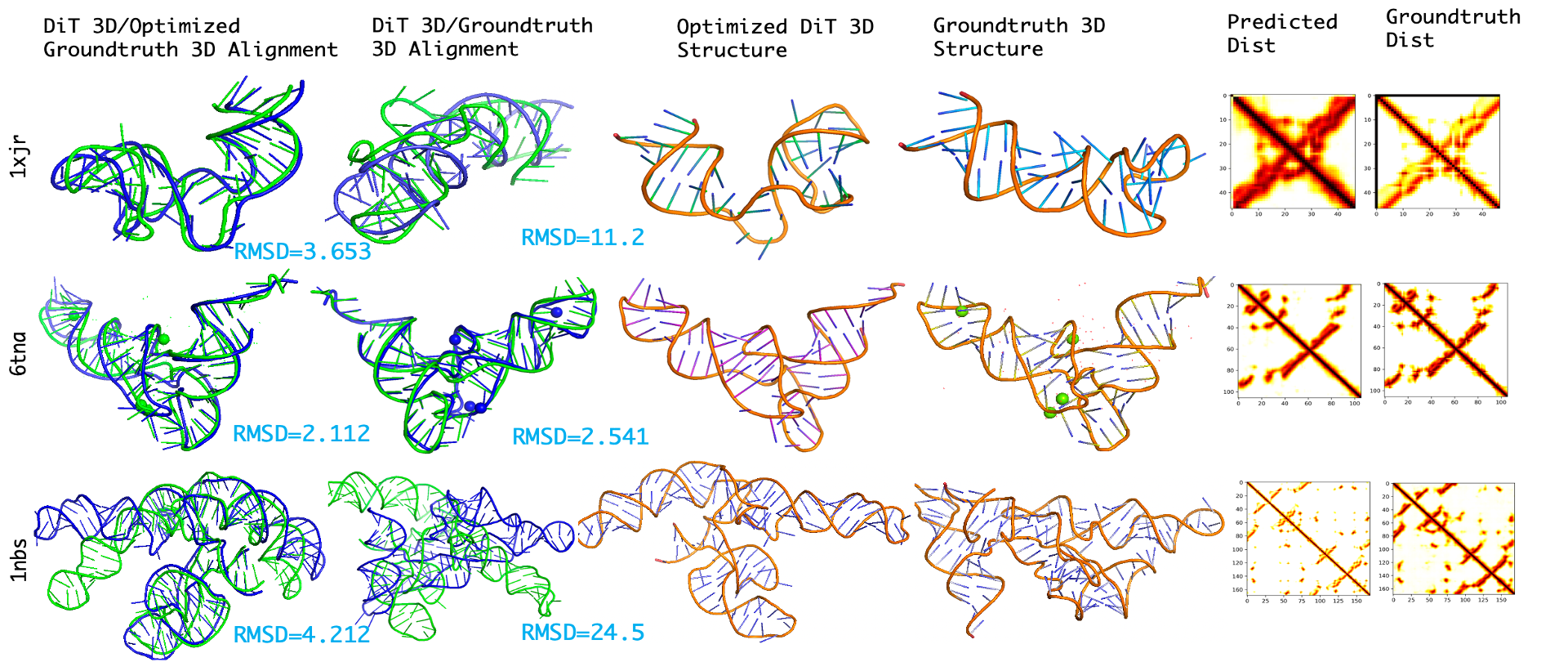}
    \caption{We reconstructed 3D-Structure of three representative RNAs to inspect the alignment results with groundtruth structures. 
    We achieved rather good RMSD on simple or medium complex structures and an acceptable RMSD regarding a complicated structure.
    Furthermore, results from the left column testified that most error derives from 3drna, optimized 3D-Structures 
    with groundtruth distance aligned properly with our previous prediction.}\label{12341234r}
    \end{center}
\end{figure*}

For convenience, we denote DiT as training a distance transformer with $1k$ distance data from scratch, and 
DiT-PS as a distance transformer with pre-training and self-training stages. TE100 stands for a size $100$ test dataset 
which contains TE80 used in \cite{sun2021rna} and removed redundancy of sequence similarity from our training dataset.
\subsection{Distance Prediction Accuracy}
Table.1 summarizes our performance on TE100 dataset when considering as a regression task. Our distance transformer with full training stages DiT-PS reaches SOTA in all five evaluation metrics. 
Interestingly, both our vanilla DiT and DiT-PS only take sequential information into account and still exceeds 
the performance of convolutional-based Unet++ trained with given secondary structures and sequences.
Unet++ trained only with sequences performs poorly and almost cannot capture helpful information. 
These results confirm the values of our trained embedding and the advantages of DiT design. Without them, it's challenging to produce 
comparable results with only sequential information.

Here, we visualized some of our prediction results in Fig.\ref{12341234t}. 
We captured almost all complicated patterns using DiT-PS in these four structures, while complex regions may be ambiguous(3jb9, 4y1m) in Unet++(Seq+SS) or vanilla DiT. 
Unet++(Seq) can only predict correctly on the diagonal line. Fig.\ref{tasd} presents our statistical results in the form of scatterplots. Dots close to the diagonal line 
refer to similar results predicted by both methods, while dots holding far from diagonal suggested ‘inconsistent’ predictions. DiT-PS outperforms all other methods,
 thus dots of MSE scatterplot almost sticked above the diagonal while dots from $R^{2}$ stayed beyond. Right part of Fig.\ref{tasd} reflects the relation between our accuracy and 
 sequence's length, noticed that the performance drops gradually when sequence's length grows. This is foreseeable as accuracy will almost decrease when structure complexity increases.

\subsection{3D-Structure Evaluation}

Distance matrixs are vital parameters in defining RNA 3D-Structures, thus we could value our task
 in the view of 3D modeling. Here, we applied 3dRNA \cite{wang20193drna} for structure folding. 
 Noticed that this method is based on optimizing an initial structure and 
 may gain errors throughout the process. Therefore the final obtained structure may be a perspective view of 
 our predicted distance but doesn't reflect the whole picture. 

 We tested on three representative structures, 1xjr, 6tna, 1nbs, 
 which shows a simple, medium, 
 and complicated structure. As shown in Fig.\ref{12341234r}, our predicted structures are well aligned in simple and medium cases with RMSD of $11.2$ and $2.5$.
  For 1nbs complicated structure, we achieved RMSD of $24.5$ which is still acceptable when predicted from raw sequences. To see that most errors indeed derive 
  from 3drna, we also include a left column to compare with the results produced by 3drna when groundtruth distance is input. We gained pretty low RMSD results regarding three different structures.

  \begin{table}
    \centering
    
    \begin{tabular}{cccccc} 
    \toprule
     \multirow{2}{*}{Methods} & \multirow{2}{*}{ACC} &
    \multicolumn{4}{c}{Long-Range Top Precision}  \\ 
           &      & L/10 & L/5  & L/2  & L/1    \\ 
    \midrule
          RNAcontact* (Seq)  & Nan & 0.48 & 0.45 & 0.40 & 0.33  
    \\   RNAcontact* (Seq + SS) & 0.92 & \textbf{0.89} & \textbf{0.88} & \textbf{0.81} & 0.66
    \\   Unet++ (Seq+SS)  & 0.88 & 0.82 & 0.79 & 0.76 & 0.55
    \\   DiT  & 0.93& 0.85 & 0.82 & 0.80 & 0.67
    \\   \textbf{DiT-PS} & \textbf{0.95} & 0.87 & 0.86 & 0.79 & \textbf{0.68} \\\bottomrule
    \end{tabular} 
    \caption{Evaluation on RNAcontact's TE80 test dataset, 
    we compared on their defined Long-Range Top 
    Precision and we also included an additional ‘ACC’ metric. 
    We achieved comparable results in several aspects to RNAcontact with 100 ensembles.
    Here * stands for an ensemble method in their paper}
    \label{2312}
    \end{table}

\subsection{RNA Contact Evluation}

We downgraded our predicted distance to compare with other existing methods in the field.
A contact is defined when the distance between two nucleotide bases $<10$ and their relative position in the sequence $>4$. 
Therefore, we can generate contact maps from our distance predictions. To calculate TopL/5 accuracy consistent with \cite{sun2021rna}, 
we select the smallest (other than diagonal line) L/5 predicted pair distance under $10$ and treat them as Top L/5 predicted contacts. 
Similarly, we can calculate Top L/10 accuracy and Top L accuracy. Table.\ref{2312} summarizes our performance on their TE80 dataset.

\begin{figure}[h]
    \begin{center}
    \includegraphics[width=0.40\textwidth]{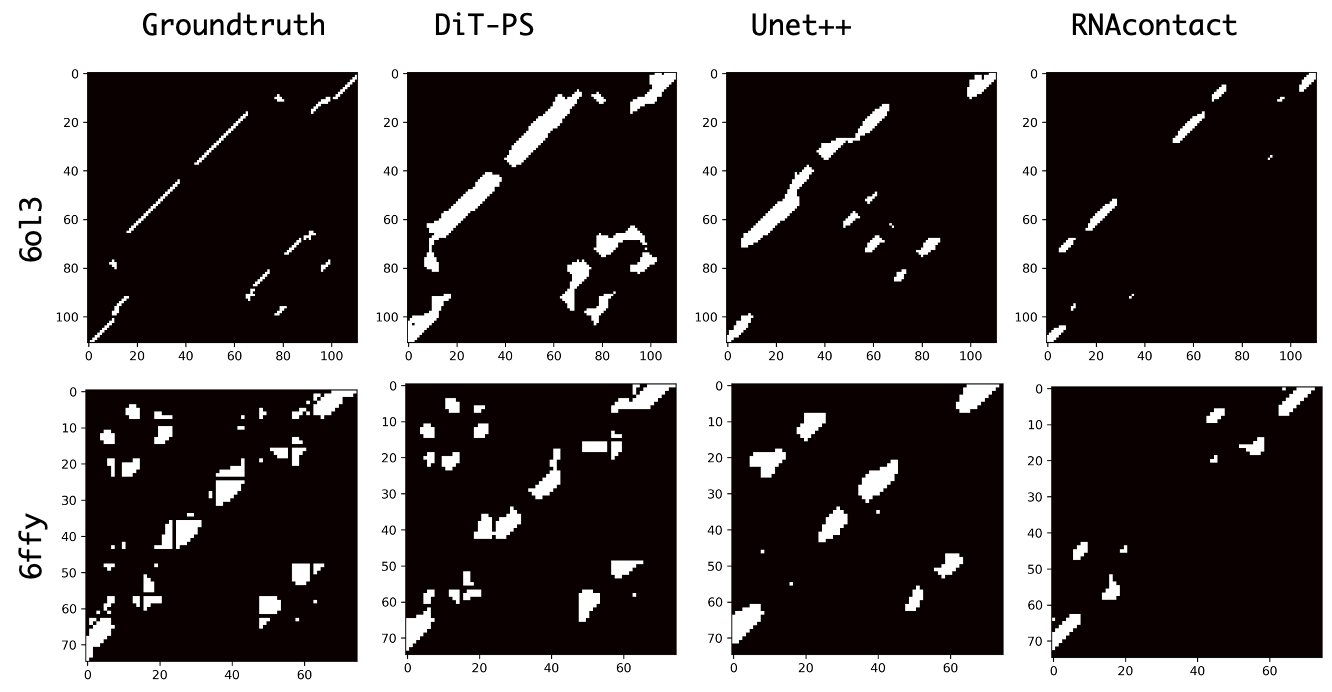}
    \caption{Visualizing RNA contact predictions of our model compared with RNAcontact method.  RNAcontact can well predict true positives while DiT performs more balanced in predicting the whole picture.}\label{cc}
    \end{center}
\end{figure}
We achieved comparable performance even downgraded to contact prediction task and obtained the best Top L precision. As reported in \cite{sun2021rna}, 
they're using aggregating RNAcontact with 100 ensembles which we denote RNAcontact* in the table to clarify.
Here we also introduce accuracy into this task as precision only takes ‘positive’ predictions into account and 
may not get the whole picture. Calculating the accuracy of RNAcontact(seq) may not be available so we leave as Nan. We also reach SOTA when considering accuracy metric.

\begin{table}
    \centering
    
    \label{1234}
    \begin{tabular}{ccccccccc}\toprule
      \multirow{2}{*}{Methods} & $\multirow{2}{*}{Enc Nums}$ 
    & $\multirow{2}{*}{Hid Size}$ & $\multirow{2}{*}{Heads}$ &  $\multirow{2}{*}{MSE}$ & $\multirow{2}{*}{MAE}$  \\ \\\toprule
  \textbf{DiT-PS} & \textbf{4} & \textbf{200} & \textbf{8} & \textbf{0.021}  & \textbf{0.124}  \\
       DiT-S & 2 & 100 & 4 & 0.039& 0.281  \\
 DiT-L & 8 & 400 & 12 & 0.047 & 0.326 \\
 DiT-H & 12 & 800 & 16 & 0.054 & 0.372 \\\bottomrule
\end{tabular}
\caption{Performance Evaluation of on the independent test Dataset TE100 of DiT variants.}
\label{1234}
\end{table}

Examples of predicted contacts of structures 6ol3 and 6ffy are shown in Fig.\ref{cc}. We can conclude that RNAcontact may achieve better precision due to its tending to predict less ‘positive’ contacts, and may
 degrade in accuracy since accuracy will consider true negatives. DiT-PS, on the other hand, is balanced in predicting the whole picture.

\subsection{Large Model Hurts Performance}

To further investigate how DiT variants perform, we tested them on TE100.
Table.\ref{1234} gives the regression error on different DiT variants 
where S,L,H behind DiT states for small, large and huge, 
we train the variants following the same training scheme as DiT-PS. 
Similar to \cite{dosovitskiy2020image},
we discovered that as transformer based model requires 
massive amount of data, DiT-L and DiT-H don't 
scale well in our small distance $1k$ problem. Increases in attention-heads don't have much impact on the performance, either.
Therefore, applying our standard DiT-PS model for further studies is better.

\section{Conclusions}
In this work, we first examine the RNA distance prediction problemand propose a general framework for inferencing only from sequential information.
Furthermore, combing with the 3drna optimizing process, we can infer RNA 3D-Structure directly.

\bibliographystyle{unsrt}  
\bibliography{references}  


\end{document}